# BIOSTIRLING-4SKA : A cost effective and efficient approach for a new generation of solar dish-Stirling plants based on storage and hybridization

*An Energy demo project for Large Scale Infrastructures*

Domingos Barbosa, Paulo André, Paulo Antunes,
Tiago Paixão, Carlos Marques
Instituto de Telecomunicações, Aveiro, Portugal
dbarbosa@av.it.pt

Arnold van Ardenne, Dion Kant
ASTRON
Dwingeloo, The Netherlands
ardenne@astron.nl

Luis Saturnino Gonzalez, Diego Rubio
GONVARRI STEEL SERVICES
Corvera. Asturias Spain
luis.saturnino@gonvarri.com

Lourdes-Montenegro, Emilio Garcia
Istituto de Astrofisica de Andaluzia
CSIC
Granada, Andaluzia, Spain
Lourdes@iaa.es

Irene Calama, Francisco Caballero
ALENER
Sevilla, Spain
Irene.calama@alener.es

Per Eskilson, Lars Gustavsson, Conny Anderson,
Jonah Lindh
CLEANERGY,
Åmål, Sweden,
per.eskilson@cleanergy.com

Reinhard Keller
MPIfiR,
Bonn, Germany
rkeller@mpifr.de

Manuel Silva Perez, Gonzalo Lobo, Valeriano Ruiz
CTAER
Sevilla, Spain

Javier Pino, Luis Valverde
University of Sevilla
Sevilla, Spain
fjp@us.es

Carsten Holze, Daniel Isaza, Rik Misseeeuw
ToughTrough GmbH (TT)
Bremen, Germany
carsten.holze@toughtrough.com

Norbert Pfanner, Alexandre Schies,
Fraunhofer ISE
Freiburg, Germany
norbert.pfanner@ise.fraunhofer.de

Jukka T. Konttinen
Tampere University of Technology,
Tampere, Finland
jukka.t.konttinen@tut.fi

The Biostirling Consortium
Zabala Innovation Consulting
Sevilla, Spain
biostirling@zabala.es

*Abstract*— The BIOSTIRLING - 4SKA (B4S) is a EU demonstration project dealing with the implementation of a cost-effective and efficient new generation of solar dish-Stirling plants based on hybridization and efficient storage at the industrial scale. The main goal of the B4S demonstration project is the generation of electric power using simultaneously solar power and gas to supply an isolated system and act as a scalable example of potential power supply for many infrastructures, including future sustainable large scientific infrastructures. B4S build an interdisciplinary approach to address reliability, maintainability and costs of this technology. [1]In April 2017, B4S successfully tested in Portugal the first world Stirling hybrid system providing about 4kW of power to a phased array of antennas, overcoming challenges in Stirling and hybridization and smartgrid technologies. B4SKA Consortium, with fourteen companies from six European countries, has performed the engineering, construction, assembly and experimental exploitation, under contract signed with the European to develop on off-grid demonstrator in Contenda (Moura) Portugal.

[1] Biostirling Project has received funding from the European Union's Seventh Framework Programme under grant agreement no 309028.

*Keywords; Concentration solar power, Stirling hybrid engines, efficiency, storage, mini-grid, large scale infrastructure, radio astronomy,*

I. INTRODUCTION

Activities toward Large science projects such as the planned Square Kilometre Array (SKA) [1] toward a next generation of Radio Astronomy telescopes, enable innovative approaches to a reduced power footprint. This is done through pushing innovative approaches to renewable energies as well as to low power computing with improved algorithms, in a domain were high density big data processing and imaging computing is mandatory. The Energy Sustainability of large-scale scientific infrastructures led to consider the impact of their carbon footprint and Power costs into the respective development path and lifetimes [6].

B4S is a demonstration project dealing with the implementation of a cost-effective and efficient new generation of solar dish-Stirling plants based on hybridization and efficient storage at the industrial scale [3]. The B4S run between June 2013 and untill April 2017. The B4S Project is based on the achievement of four targets simultaneously: to reduce costs, to increase the efficiency, to optimize the dispatchability and to increase the life-time, in order to validate a new commercial solar dish technology at demonstration scale.

Solar dish-Stirling systems have demonstrated the highest efficiency of any solar power generation system, by converting nearly 31.25% of direct normal incident solar radiation into electricity after accounting for parasitic power losses. Therefore, the solar dish-Stirling technology is anticipated to surpass parabolic troughs by producing power at more economical rates and higher efficiencies [9]. However, the aforementioned technology is not commercially exploitable to date as other Concentrated Solar Power (CSP) technologies, such as tower and parabolic solutions. This is because the current solar dish-Stirling technology still presents several limitations: high costs, limited lifetime, low system stability and reliability. Since solar dish-Stirling systems are modular, each system is a self-contained power generator, which can be assembled into plants ranging in size from kilowatts to 10MW (see Figure 1). The solar dish-Stirling plants are a hybrid solution to produce energy. This hybrid solution is a mix among different green energies, solar dish-Stirling technology as the main one, together with biomass or similar, in order to be able to produce energy 24 hours per day, which is of special interest infrastructures requiring 24/7 power availability. The initial goal of the B4S demonstration project was the generation of clean electric power using simultaneously solar power and biomass energy in the form of biogas or fuel, to supply an isolated system and serve as a scalable example of power supply for future developments and potential inclusion or consideration in large scientific infrastructures.

The SKA radiotelescope, to be installed in Africa and Australia will be once built the largest scientific infrastructure on Earth, with high and very stringent energetic demands [2,3,5,7]. Hence, it has been considered an ideal framework to have new radioastronomical concept demonstrators as reference loads for B4S, and to guide some of the specifications of the plant. Additionally, a demonstrator of one of SKA related radiotelescope technologies, installed at Moura (Portugal), has been used as a real existing demonstrator to be fed by means of B4S plant. These collaborative activities coupled to Large science projects such as the planned Square Kilometre Array [1,2,6] enable innovative approaches to decarbonization of large scale infrastructures.

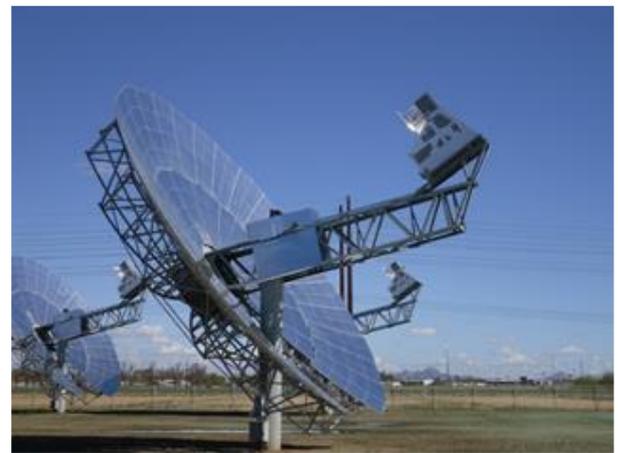

*Figure 1- Dish-Stirling engine Power plants*

The objective of BIOSTIRLING-4SKA is to research, develop and implement new solar dish-Stirling technology suitable for large-scale commercialization with the following overall goals: i) to implement a cost effective and efficient new generation of solar dish-Stirling plant based on hybridization and efficient storage at the industrial scale and ii) to study compatibility of innovative power production units with the strict Radio interference requirements of modern radiotelescopes and related radioastronomy SKA-like technologies.

II. THE BIOSTIRLING PLANT

This project in Moura developed and commissioned a prototype with an expected average power of up to 9kW. The demonstrator has served as an excellent testbed to verify the adequacy of the Biostirling energy system as a potential clean energy source for part of the SKA antennas. The hybrid solar dish-Stirling solution is a highly efficient solar and/or gas-to-electricity energy conversion unit capable of providing 24/7 electrical power. It consists of:

• A solar concentrator (Dish Unit), which concentrates the solar radiation on its focal point.

• A solar receiver placed at or in the surroundings of the concentrator focal point, which converts the concentrated solar radiation to heat by increasing the internal energy of a working fluid.

- A syngas burner, which burns syngas produced from biomass and produces sufficient heat to run a Stirling engine on the heat generated.
- A hybridization apparatus, which makes it possibly to seamlessly run the Stirling engine at continuous power on any mix of heat from the two sources above.
- The Stirling engine turns a generator from which electrical power is distributed. To be able to control the unit there has to be an engine control unit (ECU). The ECU contains a microprocessor carrying necessary control software, power supplies, sensor interfaces, safety functionality etc.

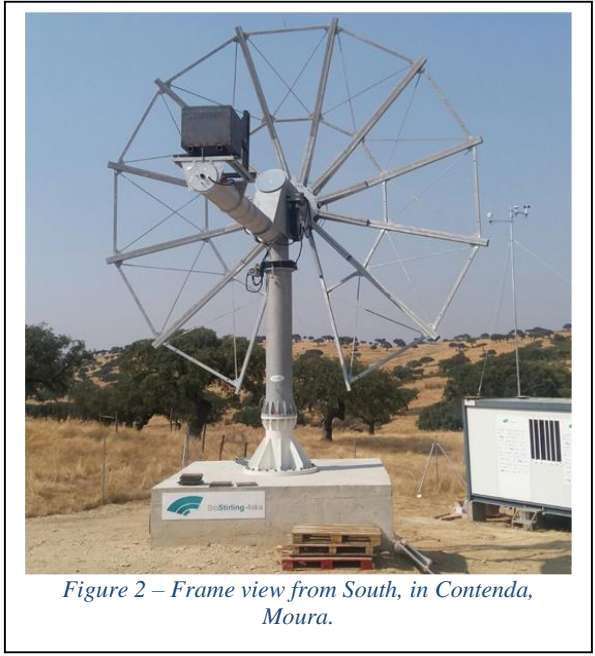

*Figure 2 – Frame view from South, in Contenda, Moura.*

The B4S prototype has been tested with the SKA demonstrator in Moura. This demonstrator consists of a set of medium-frequency antennas that are in a technology development phase and which will form part of SKA phase II. This demonstrator has been installed in the National Park of Contenda, in Moura (Southeast of Portugal), whose radiofrequency spectrum is similar to that of the deserts of South Africa and Australia where SKA will be installed.

A. *Mirrors efficiency and optical accuracy*

The efficiency of a solar concentrator is significantly affected by its optical performance. Hence, obtaining the maximum concentrated solar flux on the receiver requires a precise alignment of the mirror facets. The alignment includes the adjustment of the normal direction of the mirror surface. This orientation alignment is very important due the large influence of Direct Normal Irradiance (DNI) concentrated on the receiver. By means of canting the mirror surface, the alignment of mirror facets can be achieved. Several methods have been developed and applied for the different types of CSP collectors (i.e. parabolic trough, linear Fresnel, power tower and dish/Stirling). Basically, there are three main types of alignment: On-sun alignment, mechanical alignment and optical alignment. In order to execute the canting process, a single facet has been calibrated as a reference during sun tracking as described earlier. Following this process, the tracking mode has been deactivated and manually directing/aligning the collector with fire tower (750 m distant).

In summary, the dish structure and the optical accuracy have been verified allowing to proceeding with engine commissioning, storage integration and all related activities to performance monitoring an evaluation.

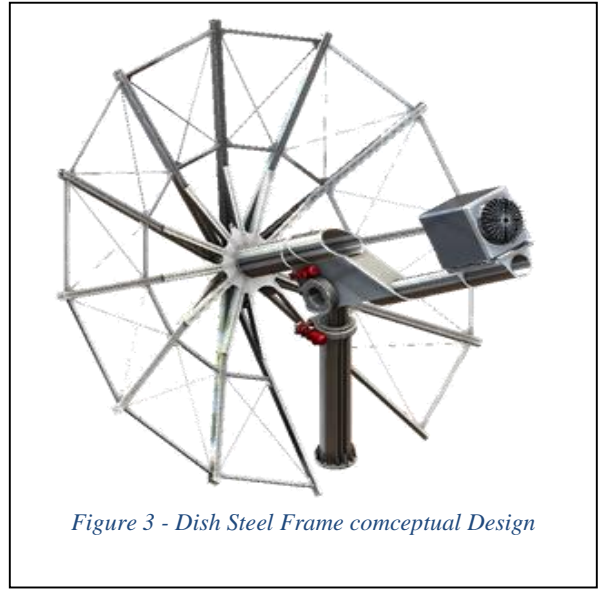

*Figure 3 - Dish Steel Frame comceptual Design*

On-site installation, integration and commissioning tasks can be summarized as follows:

- Measurement and evaluation of the static misalignment level linked to the dish main structure.
- Integration, adaption and adjustment of the steel adapters to the dish main structure.
- Compensation of misalignment by adjusting the mounting pins of the steel adapters.
- Preparation, integration and pre-canting of mirror facets.
- Laser-based evaluation of dish structure geometry.
- Geometrical deformation has been resolved after effective coordination with GMSC.
- Application of sophisticated Canting of mirror facets.
- On-site burning test: validation of optical accuracy (e.g. reflectivity, tracking accuracy, focal length and spot size).

B. *Energy Management System (EMS) Installation*

The installation of the Meter and Energy Management System (EMS) designed by ISE is performed onsite with the assistance of ALENER. The EMS box controls and measure the flux of energy coming from the public grid and also the power being produced by the plant and injected into the public Eletricidade de Portugal (EDP) grid. The rest of the boxes are meant to measure the power consumed of given by the Stirling engine, the Storage System and the antennas providing the power load.

All the equipment is also connected with communication cables. The cables used are Ethernet CAT 5 cables. It is also needed to install a switch in order to establish communication with all the meter boxes (Stirling, storage, antennas and public grid). The boxes already have a network RJ45 connector.

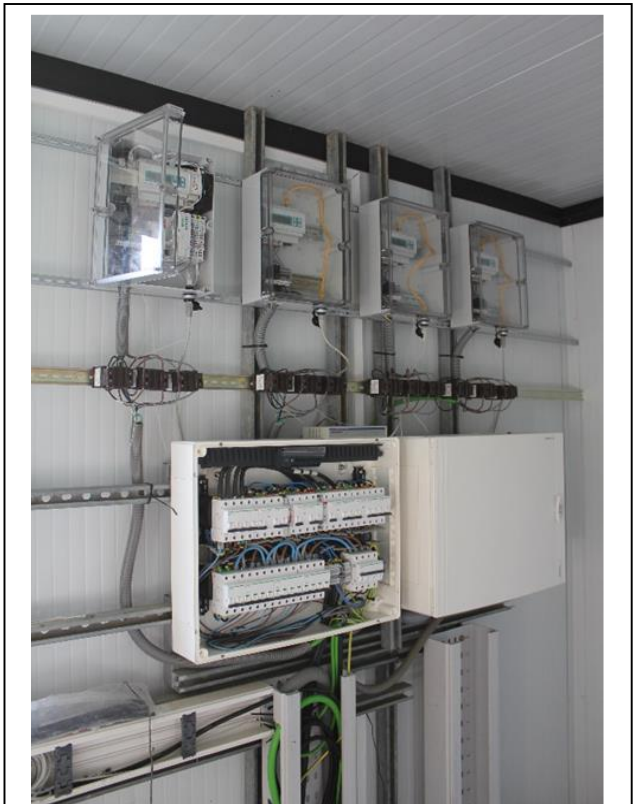

*Figure 4 - EMS installation*

## C. Storage System

The energy storage system (ESS) is located inside the site cabin. The storage system is composed of 25 modules of batteries, of 29,3 kg each, that are positioned on a 57,8 kg bank. The total weight is around 800 kg. The EES, converter and inverter system works has a UPS in order to have a backup source power in case of power loss. It has been designed and sized to supply power to tracking system, auxiliaries and SKA Demonstrator antennas in the experimental plant (when the Stirling engine fails). It can feed all the consumptions (considered) during more than 3 hours.

For its installation, two groups of cables were put out from the main switch, one going for the UPS and another for the Stirling switch.

## D. Engine Installation and Commissioning

The final design consists of 48 wickless thermosyphons oriented in two concentric circles, enabling both a tubular sun receiver and a radial gas flow through the pipes. The design included the development of a high temperature thermosyphon capable of using two different heat sources. Before sending the engine to Contenda for its commission, the prototype was deeply tested in Åmål (Sweden) with very positive results:

- The first hybrid receiver prototype was tested successfully on Cleanergy's SunBox engine in combustion only mode, generating a maximum electrical power output of 7.7kW with an electrical efficiency of 17% in January 2016.
- Different heating powers were evaluated at an inclination of 15° and all results indicated that the developed receiver will also perform well when heated only from the sun or from both sun and gas simultaneously.
- At the first test run of the unit in hybrid mode in Moura April 25th, the output power increased from combustion mode at 1,5 kW to hybrid mode producing 3 kW proving that the concept is working.

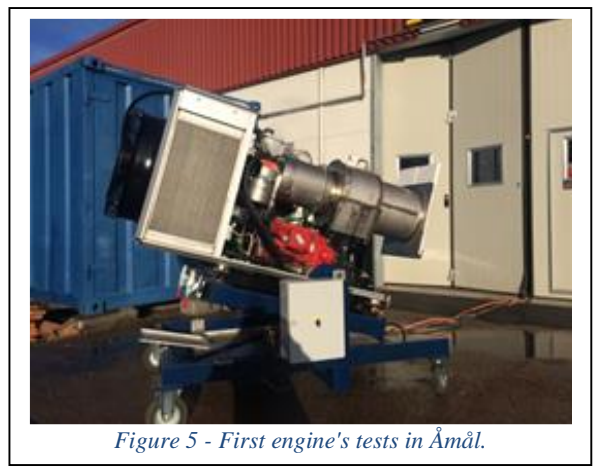

*Figure 5 - First engine's tests in Åmål.*

Before installing the Hybrid Stirling Engine, a solar version of a Stirling engine was installed in order to check the proper functioning of the system. The solar engine had been used before in Seville and was ready to run after checking and filling working gas and cooling water. No real commissioning of the engine was done since the main purpose was to check all other systems before moving on to the hybrid.

## E. Impacts & Innovation drivers

The main impacts of B4S in terms of energy-related developments are the following:

- A hybrid dish-Stirling engine that works simultaneously using solar power and gas energy. This is a very important step in order to obtain a solar-energy system that can work 24x7.

- Our solar dish-Stirling prototype has an electrical power of 10Kw (1-9 kW in gas mode, 2-10 kW in solar mode), what can provide energy supply isolated systems and serve as a scalable example of power supply for future developments of larger infrastructures.

- Several important improvements on the structure of the dish by reducing its weight and improve tracking precision.

- Regarding optical parameters, the concentrator has been evaluated with an average reflectivity of more than 95,5% and an average optical slope error of 1.7 mrad at maximum.

- Assuming 20% efficiency from gas chemical energy to power, the chemical energy demand is 250 kWh.

- A new energy storage system able to work simultaneously with the hybrid technology, hence able to avoid non-desirable energy peaks.

- An innovative control system that achieves a highly reliable and fully renewable hybrid solution.

## III. THE SKA

### A. Radioastronomical Large Scale Infrastructures

Large sensor-based science infrastructures for radio astronomy like the SKA will be among the most intensive data-driven projects in the world, facing very high demanding computation, storage, management, and above all power demands. The geographically wide distribution of the SKA and and its associated processing requirements in the form of tailored High Performance Computing (HPC) facilities, require a Greener approach towards the Information and Communications Technologies (ICT) adopted for the data processing to enable operational compliance to strict power budgets. Addressing both the reduction of electricity costs and the generation and management of electricity at system level is paramount to avoid future inefficiencies and higher costs, while enabling fulfillments of Science Cases like Transient observations or other Virtual Observatory (VO) triggered observations whose operational modes may produce sudden peak power loads. In particular, since the (power hungry) data processing location is conditioned by the experiment, and not by the computational facilities, it results in far from optimal efficiency, higher capital expenditure (CAPEX) and higher OPEX. As an example, VO triggered events will require precise metering, operating under the constraints of Power forecast budget, while considering power buffer allocations for these unexpected, yet extremely relevant astronomical events. Management of these operational modes require power forecast at subsystem level and power buffers, eventually with reconfiguration, disconnection or graceful operational downgrade of some other telescope components. Hence, at system planning, we must consider a combination of low power computing, efficient data storage, local data services, Smart Grid power management, and potential inclusion of other non-grid sources like Renewable Energies in the form of heterogeneous system mix, and a heterogeneous power mix at provision level.

### B. New Concepts

As an example connected to B4S, astronomical infrastructures are usually built in remote locations, turning power supply a potential sizeable fraction of capital and investment costs. Distributed facilities over hundreds of Kms like the Square Kilometre Array (SKA), to be installed in Africa and Australia deserts present serious and very stringent energetic demands. SKA, is a project to become the largest science infrastructure of XXI century, has been considered an ideal example to have as a reference for B4S , to guide some of the specifications of a demo plant. Additionally, a demonstrator of one of SKA Advanced Instrumentation technologies installed at Moura (Portugal), has been used as a reference isolated system to be fed by the B4S demo plant..

The expected average power usage of the whole SKA will be between 50-100 MW, but over an extended location (up to 3000 Km diameter), with many different nodes, and sparse occupation of that terrain beyond the central core. Since SKA will scan the sky continuously, it will not present strong power peaks and power fluctuations, requires a smooth consumption profile. Energy generation at a continental scale for this facility, with different load profiles at different locations, means that modular power generators are needed, presenting an ideal scenario for development of innovative solutions with its own degree of customization and grid connectivity. Another consideration is that SKA, by definition, requires 24/7 observation operations. Another consideration is that SKA, as a radiotelescope, can observe the sky 24/7, so power consumption should also be maintained night and day, and free of radiointerferences, given the extremely faint signals to be observed. Because of these technical requirements, the power supply considerations for SKA present an opportunity for inclusion of smart, more efficient and low-carbon technologies. It is at this point where B4S and SKA are connected.

Apparatus for the radio measurements include 4 tiles of an MFAA prototype based on Electronic Multi Beam Radio Astronomy ConcEpt (EMBRACE) [4,5]. EMBRACE demonstrates the design readiness of the phased array technology for the Square Kilometre Array (SKA). There are two major EMBRACE stations, one in Nançay, France, and the other one at the Westerbork Synthesis Radio Telescope (WSRT) in the Netherlands [8]. However, for deployment in South Africa EMBRACE requires qualifying for environmental factors and compatibility studies with the inclusion of sustainable energy sources. The former is also addressed with installation of EMBRACE modules similar to those in Contenda in the SKA South Africa Karoo site. Renewable's inclusion in a system with new SKA technologies is therefore performed as a world's first in Contenda. On the concept level the reliance on a "software telescope" such as these phased array concepts, reduces the number of stations as the collecting area per station is larger. Interesting enough this reduces the requirements of the central processing units and hence of the corresponding power needs. As now, the station requires a probably higher power as a result of the pre-processing and aggregate needs of the receivers, this invites the needs for a station level power solution. At the same time, this may reduce the costly copper requirement to the central processing when using fiber optic interconnect-and network solutions, which simultaneously increase radio interference and lightning (if any) effects.

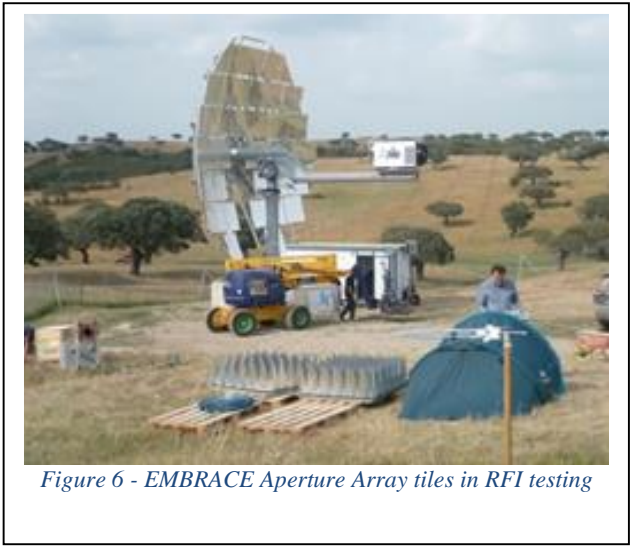

*Figure 6 - EMBRACE Aperture Array tiles in RFI testing*

C. Commissioning and RFI testing : Demo Aperture Array tiles

The Bio-Stirling Power Plant was connected to the SKA Mid Frequency Aperture Array (MFAA) demonstrator and started working in simulated real conditions as the only electric supplier of the radio-astronomical system as a whole. MFAA tiles require a stringent steady state power supply, besides a good control of Radio Frequency Interference (RFI) levels, in agreement with SKA RFI standards. Deploying B4SKA technology in future to deliver energy to SKA telescopes results in tough RFI emission requirements on the energy generating equipment within a short distance to the SKA stations. For Assembly, Integration and Verification operations (AIV), ASTRON provided the demonstrator including 4 antennas tiles, 1 antenna beam former, protection huts and processing unit, electrical circuitry and control. ASTRON was also responsible for its assembly and integration, while IT and CSIC were responsible for the verification tasks (Figure 6).

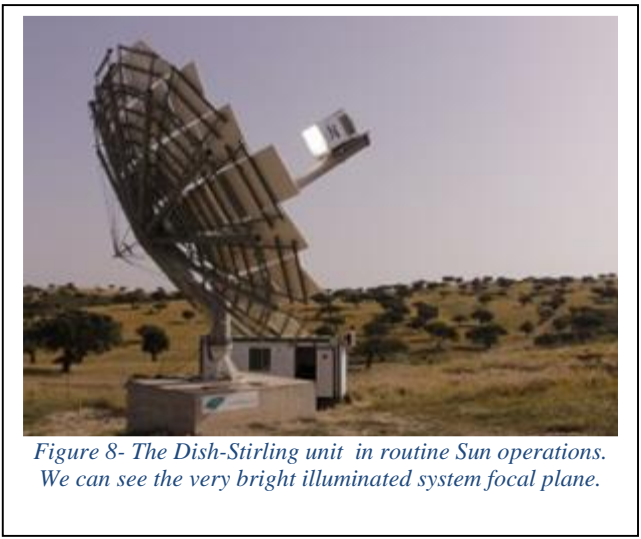

*Figure 8- The Dish-Stirling unit in routine Sun operations. We can see the very bright illuminated system focal plane.*

IV. CONCLUSIONS

The B4S project is a demo project, with a high Technology Readiness Level (TRL). B4S aimed to push forward the solar dish–Stirling technology towards full dispatchability in the smart.grid field or power islands, reduce cost and increase efficiency and life-time of these concentrating solar power systems. The concepts and implemented technical solutions were considered for application in the large scale project SKA. B4S achieved the final implementation of one integrated unit dish-reflector with hybrid receiver and Stirling engine and associated PCU+electrochemical storage (batteries) in order to validate a new commercial solar dish technology at demonstration scale. The Bio-Stirling-4SKA (B4SKA) Consortium, with fourteen companies from six European countries, led by Gonvarry (Spain) has performed the engineering, construction, assembly and experimental exploitation, under contract with the European Commission for the Seventh Framework Program (7FP) of a solar concentration system powering a set of demonstration astronomical EMBRACE - MFAA antennas installed in Contenda Forest (Moura, Portugal). The power provision developed a cost effective and efficient new generation of solar dish-Stirling plant based on hybridization and efficient storage at the industrial scale,

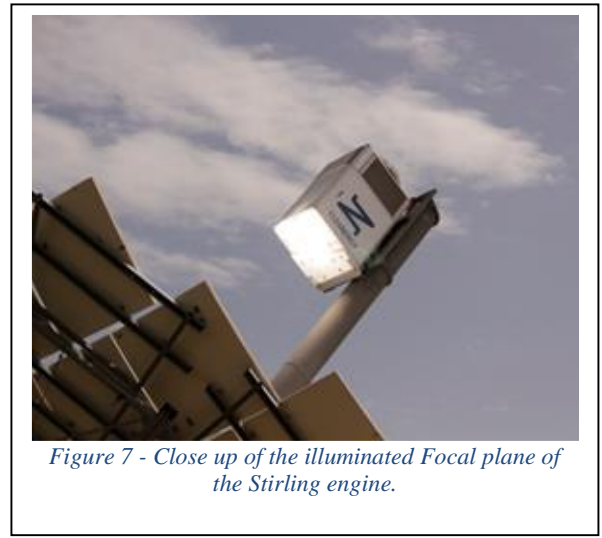

*Figure 7 - Close up of the illuminated Focal plane of the Stirling engine.*

combining solar concentration, gas and batteries for night operations.

B4S has driven innovation in the following areas:

- Hybrid Stirling motor.
- Development of passive cooling system and smart load sequencing to minimise demand side energy requirements.
- Generation of serendipitous radio frequency interference shielding, interfacing with grid supplies and land use.
- Distribution and reticulation design for price/reliability trade-offs
- It has improved the conversion efficiency above 24%, through the new design of the parabolic concentrator, innovations in the coating materials to replace the conventional mirrors and increase reflectivity, modelling and optimisation of operation, and definition of a control system to implement all aforementioned improvements.

The new plant achieved "first light" on 25th of April 2017, Freedom day in Portugal, by providing about 4kw of power to a set of EMBRACE – MFAA antenna tiles (Figures 7,8). Astronomical institutes were responsible for requirements, RFI measurements and deployment planning of the system. Besides the testing of dispatchability and compatibility with a radioastronomical system, B4S actually implemented the first world example of an hybrid concentrator engine, opening new avenues for further innovations of green autonomous radioastronomical systems with greater economic impact.

Powering with green concepts new radioastronomical Aperture Array stations in the hugely beneficial conditions at the "SKA" sites, appears therefore an obvious choice advantageously positioning the SKA as a "green" telescope while reducing the operational costs and its carbon footprint. Such were key arguments to advance innovative solutions like explored in the project Biostirling for SKA ("B4S").

ACKNOWLEDGMENTS

This project has received funding from the European Union's Seventh Framework Programme (FP7) for research, technological development and demonstration under grant agreement no 309028. DB acknowledges support from ENGAGE SKA Research Infrastructure, POCI-01-0145-FEDER-022217, funded by Programa Operacional Competitividade e Internacionalização (COMPETE 2020) and FCT, Portugal. L.V.M. acknowledges as well support from the grant AYA2015-65973-C3-1-R (MINECO/FEDER, UE)